\documentstyle[]{article}

\begin{document}
\author{Gary Oas \\  Education Program for Gifted Youth \\  Stanford University}
\title{On the Use of Relativistic Mass in Various Published Works}
\maketitle
\begin{abstract}
A lengthy bibliography of books referring to special and/or general relativity is 
provided to give a background for discussions on the historical use of the concept
of relativistic mass. 
\end{abstract}


\section{Introduction}  

The primary purpose of this paper is to provide a bibliographic base to the discussion
of relativistic mass (RM). In a separate paper \cite{oas}, arguments dissuading the use of the 
concept of a velocity dependent mass are put forth and the need for such arguments
is based on the continued widespread use of the concept. This article
provides a reference list to the specific works listed in the survey. 

The 634 works are separated into categories as presented in \cite{oas}: textbooks devoted
to special and/or general relativity; popularizations of relativity, physics, or modern physics;
introductory and modern physics textbooks; and a miscellany of other works. In certain
sections, in addition to the categorization of the use of RM, other facts of interest have been included.
For example, in listing relativity texts the signature of the metric (if appropriate) is given as
well as whether naturalized units (c=1) are employed. 
Additional commentary on several of the works is included here, most in relation to the use of RM.
A summary of the number of works employing relativistic mass and those that do not is given
at the end, these numbers yielding the plots in \cite{oas}.

A brief comment about the manner in the classification of these works is in order. Whether
or not a work employs relativistic mass is not always an objective choice. There are
instances where an author may introduce the concept and then rail against it, while others 
may not explicitly introduce it yet allude to its validity, and in some small circumstances
(found in introductory textbooks) make explicit contradictory statements.
Thus in several instances it has been the judgment of this author as to whether it is employed or not. Those instances
have, for the most part, been noted in the comments. The basis for judgment has been
from the perspective of a first time learner; if the presentation leaves an impression that
this concept is valid (regardless of its fashionability) then it has been listed as having
introduced the concept. Some may disagree with the decision on some of these references and may
be up for debate, however the vast majority or works take a clear-cut stance one way or the other.  


The works here provide an extensive, but not exhaustive, base to evaluate the use of the
concept of relativistic mass throughout the past century.
This reference is a work in progress, it will be updated as new works are published as well
as other, older, works are examined. The most up-to-date version of the bibliography will be made available on the internet\cite{rmweb}.
This bibliography will also be made available as both a spreadsheet and in Bibtex format. 
Therein is listed works that have not been examined.


\section{Textbooks Devoted to Special and/or General Relativity (SGR)}

There are 100 books that have been examined within this category. They are separated
into two sections, those that do not employ RM and a second section listing those that do employ
the concept. Of these works 63 introduce RM and there is not observed any significant historical trend away from  or towards the concept. The number of texts published that do not use the concept has remained
relatively constant over the past 35 years. 

In the comments that follow each reference the signature of the metric is given as well as
whether naturalized units (setting c = 1) are used. For the signature there are three
possibilities; [W] are ones where the signature is (+1, -1, -1 ,1); [E] where 
(-1, +1, +1, +1); and [M] which refers to Minkowski's complex time coordinate (i.e. $ct\rightarrow ict$)
found in some older texts (which leads to a metric of signature [E]). 
As for units; [N] refers to the use of naturalized units and [S]
refers to standard units where explicit reference to $c=3 \times 10^8 m/s$ is continually made.

\subsection{SGR Works that Do Not Employ RM}



\subsubsection{Footnotes}
\begin{itemize}
\item \cite{SGRn31} 

He uses the term "rest mass" continually in the section on SR. We shies away from saying that mass increases with speed but in the tensor chapter he does say it twice. 

The problem here is that in using naturalized units, c = 1, energy appears to be {\it exactly}  the same as mass, $E=m$. This can be deceptive, see Okun's article (``second argument")\cite{okun2}. Uses terms interchangeably.
\end{itemize}

\subsection{SGR Works that Employ RM}
This section lists those works that employ the concept of relativistic mass. 


\subsubsection{SGRy footnotes}

\begin{itemize}

\item \cite{SGRy23}, pg73:

``Most textbooks on relativity work with the relativistic energy $E = mc^2$ and rest mass $m$ rather than with the relativistic mass, a velocity-dependent quantity, $m(v)=\gamma v$. We shall do the same from here on to avoid possible confusion. Mass means rest mass $m$, just as in Newton's world."

\item \cite{SGRy58}, pg 169:

``Since $m*$ depends on $u$ through $\gamma$, it is customary to say that ``mass depends on velocity." But it must be remembered that this $m*$ is the {\it relative} mass; the {\it proper} mass does not depend on velocity, and the ``dependence on velocity" is introduced merely for the sake of obtaining formal agreement between the equations of relativity and those of Newtonian mechanics."

\end{itemize}


\section{Popularizations of Relativity, Physics, or Science (PS)}

	Of the 105 works examined that present a popularized, pedagogically basic, introduction
to relativity or the concepts of modern physics, only 18 were found not to utilize RM. A vast majority rely on this concept and the numbers have been growing. Since the publication of Stephen Hawking's bestseller ``A Brief History of Time"\cite{PSy43}, the number of books published presenting advanced concepts of physics to the general public has exploded (however there is recent evidence
that this trend may be reversing\cite{sciencearticle}). 

More and more, prominent physicists
have been lured into this lucrative market. In addition, non-physicists ranging from experts
in different fields to science journalists to those with no credentials whatsoever have been publishing in this vein. In fact, in the 
period 2000 to the present, in this category 40 books were examined and of these
exactly half were written by non-experts. 
The post-1999 works below list whether the author is an expert, [P] (that is, has a Phd.
in physics or astrophysics, publish original research, and/or is an instructor) or not, [NP]. 
As many of these non-expert writers rely upon older popularizations
or upon conversations with experts, it is no wonder why the number of works utilizing RM continues to grow. 

\subsection{PS Works that Do Not Employ RM}


\subsubsection{PSn footnotes}

\begin{itemize}
\item \cite{PSn3}
Ok with regards to RM, but on pg. 32 says that from $E=mc^2$, $1 kg = 3 x 10^{16} J$ and on pg. 56 says that when at rest the inside of the square root is zero when it should be 1.

\end{itemize}

\subsection{PS Works that Employ RM}



\subsubsection{PSy footnotes}

\begin{itemize}
\item \cite{PSy1}
On page 82 invokes the "Law of Conservation of Mass" 
(Lavoisier's law) to derive relativstic mass from an inelastic collision viewed in two frames.
Derivation is on pg 83.

On page 84
"We are in a sense placing rather too much emphasis on the concept of mass here. This is understandable since it plays such a central role in classical mechanics. In relativity there are several ways in which mass and inertia can be defined. Our use of 'relativistic mass'...makes the law of conservation of energy and mass equivalent statements. Once we have developed a proper relativistic concept of energy we will be able to tackle dynamic problems simply by considering the conservation of energy and momentum (and these too [sic] are closely related, as we shall see when we consider a spacetime representation in the next chapter).
\\
Page 161 end  Now goes on and does it properly
\\
``In section 2.12 we assumed a relativistic momentum of the form p = m(v)v and used this to look for an expression for the relativistic mass m(v) such that the conservation laws for mass and energy would hold in an interaction. The result was that  $m(v) = \gamma m_0$ and $p = \gamma m_0 v$ ($m_0$ is rest mass). The expression for momentum is simply rest mass multiplied by $\gamma v$ which we now recognise as the first three components of the 4-velocity."

\item \cite{PSy2}
He says there is no difference in which way you look
(rel mass or not). Explains that particle physicists do not use RM.

Quote  pg 36:
``In this approach. M is termed the {\it rest mass} of the particle. This has the advantage of explaining {\it why} Newtonian momentum is wrong -- Newton did not know that mass was not constant."

\item \cite{PSy17}
Pg 139, an exceptionally misleading quote: \\
``The answer is no, you can't get younger. But you can age more slowly than a friend (or twin) traveling at a slower speed. On the other hand, you pay a price: You also might temporarily get more massive in the process."

\item  \cite{PSy52}   
Quote from page 66:\\
``For example, Einstein could show that the mass of an object increased the faster it moved. (Its mass would in fact become infinite if you hit the speed of light - which is impossible, which proves the unattainability of the speed of light.)
This meant that the energy of motion was somehow being transformed into
increasing the mass of the object. {\it Thus, matter and energy are interchangeable!}"

\item  \cite{PSy40}  
pg 52 (whole first pp)  (Also incorrect use of "Einstein's equation") \\

From the viewpoint of the concepts we have emphasized in this chapter, Einstein's equation gives us the most concrete explanation for the central fact that nothing can travel faster than light speed. You may have wondered, for instance, why we can't take some object, a muon say, that an accelerator has boosted up to 667 million miles per hour -- 99.5 percent of light speed -- and "push it a bit harder," getting it to 99.9 percent of light speed, and then "{\it really} push it harder" impelling it to cross the light-speed barrier. Einstein's formula explains why such efforts will never succeed. The faster something moves the more energy it has and from Einstein's formula we see that the more energy something has the more massive it becomes. Muons traveling at 99.9 percent of light speed, for example, weigh a lot more than their stationary cousins. In fact, they are about 22 times as heavy--literally. (The masses recorded in Table 1.1 are for particles at rest.) But the more massive an object is, the harder it is to increase its speed. Pushing a child on a bicycle is one thing, pushing a Mack truck is quote another. So, as a muon moves more quickly it gets ever more difficult to further increase its speed. At 99.999 percent of light speed the mass of a muon has increased by a factor of 224; at 99.99999999 percent of light speed it has increased by a factor of more than 70,000. Since the mass of the muon increases without limit as its speed approaches that of light, it would require a push with an {\it infinite} amount of energy to reach or to cross the light barrier. This, of course, is impossible and hence absolutely nothing can travel faster than the speed of light.

\item \cite{PSy73}
pg 212  (this comes after correctly describing spacetime via diagrams and KE leads
to fundamental limit.
``Whether or not to speak of velocity-dependent mass is largely a matter of taste. Although it is currently unfashionable to do so, Einstein did and we shall as well."
As discussed in \cite{oas}, this is a factually inaccurate statement.

\end{itemize}

\section{Miscellaneous Works (Misc)}

Those 119 books that did not fall under any of the other categories have been lumped into
a miscellaneous category. Thus, a wide range of books are reported here. The general type of book is listed but subcategories have not been tabulated. The general type of books are referred to as:
OT, other textbooks; APT, advanced physics texts (not SGRT); Phil, philosophical works, including those of a religious nature; Hist, historical accounts; REF, reference books, such as encyclopedias and 
dictionary of physics; and SF, science fiction books.

Of these works only 14 did not introduce the concept of relativistic mass. However, this should
not point to the general disparity between the two categories as the survey conducted for this
category was biased. A large number of works were located through searches (web search engines and library catalogues) for relativistic mass. Thus it can be noted that the number of works utilizing RM is growing but no firm conclusion can be made about those that do not. It is far more difficult to search for non-occurrences of a concept. In addition, the use of internet search engines and online book stores have tended to return more modern books, thus the distribution is weighted towards the current era. Thus this category, unlike the others reported here, should be taken with a grain of salt, there are
a large number utilizing it these days but the general historical trend requires more research to be conclusive.

\subsection{Misc Works that Do Not Employ RM}



\subsection{Misc Works that Employ RM}



\subsubsection{MISCy footnotes}

\begin{itemize}

\item \cite{MISCy1}
Quote pg 17: 

``The light beam, he now understood, travelled at its constant speed while matter became smaller and heavier the farther he got from it, the more slowly time passed. 

\item \cite{MISCy8}
Quote: 

``The addition of the relativistic mass to the overall mass of Mercury produces a very small acceleration to the orbital motion of the planet. "

\item \cite{MISCy33}
Quote: 

 ``beam of particles faster than ever before by using an alternating electric field whose frequency could be adjusted for relativistic mass changes."
 
\item \cite{MISCy70}
Quote pg 222: \\

``Finally in a law like the relativistic mass law, $m = m_0/\sqrt{1-{v^2\over c^2}}$, we have neither a summary of observations nor a tendency statement, but a frame proposition expressing a ``grammatical" rule for the use of ``m", the concept of relativistic mass."

\end{itemize}


\section{Introductory Physics and Modern Physics Textbooks (IPT)}

This category of these works is the most significant reported here and more effort has
been made to classify each textbook edition. In all 315 editions of textbooks have been
examined with the general result that a majority utilize RM (223 versus 92). However, as
reported in \cite{oas}, the modern trend has been one of moving away from this concept.
This is hypothesized to result from scrutiny of the literature on physics education research
in writing introductory textbooks. Generally, one does not refer to this body of research when
writing an advanced text or popularization of physics. It is interesting to note that this
trend creates a widening gap between the two viewpoints, leading to significant inconsistencies in what is put forth. I have found that students arriving into my course on relativity have deep 
preconceptions of relativistic mass and often point to widely read popularizations
penned by some of the most prominent physicists today. To have them unlearn this
concept requires significant convincing (and time) and presents a sizable obstacle to 
an understanding of the modern geometrical formulation of relativity. Thus, those who
introduce this concept as a fact of nature are doing a disservice to those that want to go on
to become practicing relativists.


\subsection{Category A: Conceptual Introductory Physics Textbooks}

The following are references that fall under the classification of ``conceptual" introductions
to physics. This corresponds to category A as designated by the College Board. A summary and historical trends is provided in \cite{oas} and will not be repeated here. There may be slight
differences in the numbers as some works have been reclassified. 

The use of relativistic mass is given for each edition, as it sometimes changes during the
history of a textbook. Editions which utilize RM are denoted by bold years enclosed in square brackets, i.e. [{\bf 1999}]. Those editions that do not not emphasized and are enclosed in parenthesis, (1999). There are some
editions that have not been examined, these edition years are in italics, ({\it 1999}). And lastly, a small number of editions give contradictory statements on the status of relativistic mass, these are noted in the comments and listed as such, [{\bf 1999*}].
As before, relevant quotes and comments are given at the end of the listing.

\subsubsection{IPMa, footnotes}

\begin{itemize}
\item \cite{IMPa18} 2nd edition: 

-pg 128 
``The conclusion is that {\it momentum conservation can only hold if a fast-moving body appears to have more mass than a slow moving one.}"

-pg 129
``At speeds approaching that of light, the body becomes more and more massive. Then the increase in momentum involves a large change in mass and a small change in velocity.

This property of mass also assures us that Einstein's requirement that nothing move faster than light will never be violated by a material object. As a body approaches the speed of light, the application of a force in the direction of motion will only make it more massive and not change its speed appreciably."

\item \cite{IMPa18} 3rd edition:

-pg 1313 footnote
``Some relativity texts reserve the term {\it mass} for rest mass alone and put the $\gamma$ in the definition of momentum. This is simply a semantic choice."

\item \cite{IMPa9} 
Following is a quote from the 9th edition, previous editions had no such disclaimer.

``Classically, the particles behave as if their masses increase with speed. Einstein initially favored this interpretation, and later changed his mind to keep mass a constant, a property of matter that is the same in all frames of reference. So it is $\gamma$ that changes with speed, not mass.''
\end{itemize}

\subsection{Category B: Algebra-based Introductory Physics Textbooks}

These textbooks fall under category B as specified by College Board; introductory
textbooks that do not assume knowledge of calculus.

\subsubsection{IMPb footnotes}

\begin{itemize}
\item \cite{IMPb16} 6th edition:
 Clarifies his use, suggests many don't like it and to be careful with it. But still refers to it on footnote on page 744. 

\end{itemize}

\subsection{Category C: Calculus-based Introductory Physics Textbooks}

These textbooks fall under category C, calculus based introductory textbooks.

\subsubsection{IMPc footnotes}

\begin{itemize}
\item \cite{IMPc13}

-pg 15-1 pp 3
``For those who want to learn just enough about it so they can solve problems, that is all there is to the theory of relativity -- it just changes Newton's laws by introducing a correction factor to the mass."

Also see 15-9, 15-10 and 11, 16-1,  and 16-4

\item  \cite{IMPc36}
T.R. Sandin is now a contributing author (editions 10,11). As he is a vocal proponent of the use of RM, it is not too surprising that its use creeps back in these later editions.
Added sentence in 10th ed. Introduces and say there are strong opinions on both sides.

``Also with relativistic mass, the famous equation $E = mc^2$ can be applied to all types of energy, not just most types."

\item  \cite{IMPc37} 
This popular textbook has changed its position twice on the matter. And in transition, contradicting statements are made in the 3rd and 4th edition.

{\bf 3rd edition}

Pg 1124: Relativistic mass is introduced and the explicit formula presented.

Pg 1128: ``Finally, note that since the mass m of a particle is independent of its motion, $m$ must have the same value in all reference frames."

{\bf 4th edition}

Pg 1175: ``Finally, note that since the mass m of a particle is independent of its motion, $m$ must have the same value in all reference frames."

Pg 1177 ``It follows that mass varies with speed (relative to the observer). We must therefore distinguish between {\bf rest} mass, $m_0$, which is the mass measured by an observer at rest relative to the particle (and at the same location), and the mass measured in real experiments."

By 5th edition this later terminology is removed, (but still states that light has mass).

\item \cite{IMPc41}
Page 201, there is a discussion of energy and mass wherein $E_0=mc^2$ is used and continual discussion of rest energy is used. Here a good statement of the relation between mass and energy is made.

`` According to Equation 7-7, a particle or system of mass $m$ has ``rest" energy $mc^2$. This energy is intrinsic to the particle."

However, roughly 130 pages later, on page R-12, relativistic mass is introduced in relation to relativistic momentum.

``Equation R-10 is sometimes written $p=m_rv$, where $m_r$ is called the relativistic mass"...
``Relativistic mass and momentum are discussed further in Chapter 39."

\end{itemize}

\subsection{Category D: Modern Physics Textbooks}

Here are listed modern physics textbooks intended for a first time exposure to
the ideas of modern physics.

\subsubsection{IMPd footnotes}

\begin{itemize}

\item \cite{IMPd3}
In the latest, 6th, edition of this long running text the author expresses his dislike for the concept of relativistic mass. It is interesting that it took so long for this view to be expressed in print.

\item \cite{IMPd15}
States they are different approaches and equally valid.

``In any case it is necessary to check the final physical laws, no matter by what process they are arrived at, against the actual properties exhibited by nature. By this criterion either of the above expressions for the dynamical laws must be considered to be experimentally verified, but of course the criteria of simplicity and aesthetics speak strongly for the 4-vector form."

\end{itemize}

\section{A Quick Summary}

	The historical trends on the above works are presented and discussed in \cite{oas}, they will not be repeated here. However a quick summary of the use of RM in these works is provided in the following table. The results may differ slightly than in \cite{oas} as some works have been reclassified (and more refinement is still in order). 
	
	\begin{center}  \label{table1}
\begin{tabular}{|l||c|c|c|c|c|c|c|c||c|}
\hline
Category & {\small pre-1970} & {\small '70-74} & {\small '75-79} & {\small '80-84} &{\small '85-89} &{\small '90-94}& {\small '95-99} &{\small '00-05} & Total\\ \hline\hline
SGR no & 6&3&4&5&3&8&3&5&37\\ \hline
SGR yes &27 &5&4&0&3&8&8&8&63\\ \hline \hline
PS no &2 &0&2&1&3&5&0&6&19 \\ \hline
PS yes &5 &1&1&5&7&11&21&35&86 \\ \hline \hline
MISC no & 0&1&1&0&1&4&1&5&13\\ \hline
MISC yes &4 &1&1&4&7&14&28&45&104\\ \hline \hline
IMPa no &0 &1&0&0&0&2&5&8&16\\ \hline
IMPa  yes & 2&7&5&5&6&6&4&5&40\\ \hline \hline
IMPb no &2 &1&1&2&4&6&4&9&29\\ \hline
IMPb  yes &17 &8&12&8&14&7&7&5&78\\ \hline \hline
IMPc no &0 &0&1&3&3&4&8&10&29\\ \hline
IMPc  yes &20 &8&5&12&7&4&9&3&68\\ \hline \hline
IMPd no &0 &1&3&1&2&0&5&6&18\\ \hline
IMPd  yes & 17&5&3&5&3&1&2&1&37\\ \hline \hline
ALL no & 10&7&12&12&16&29&26&49&158\\ \hline
ALL yes &92&35&31&39&47&50&69&91&476\\ \hline
\end{tabular} 
\end{center}

\vspace{0.5 in}

{\bf Acknowledgments:} I would like to thank Paul Dimitre and Wendy Stallings for help in locating certain hard-to-find textbooks, in addition to my friends and colleagues who allowed my to rifle through their collections.  I would also like to thank the Education Program for Gifted Youth for support during this time.

\end{document}